\documentclass{IEEEtran4PSCC}

\usepackage{cite}
\usepackage{bm}
\usepackage{amsmath,amssymb,amsfonts}
\usepackage{algorithmic}
\usepackage{textcomp}
\usepackage{xcolor}

\usepackage{tabularx}

\usepackage{mdwlist}
\usepackage[ruled,vlined,shortend, linesnumbered]{algorithm2e} 
\usepackage[]{algorithmic} 
\usepackage{booktabs}
\usepackage{color}

\usepackage{amsthm,amssymb}
\usepackage{amsmath}
\usepackage[]{caption}
\usepackage{subcaption}
\usepackage{dblfloatfix} 

\usepackage{booktabs}

\usepackage{multirow}
\usepackage{mathrsfs}
\usepackage{amsbsy}
\usepackage[mathscr]{eucal} 
\usepackage[flushleft]{threeparttable}
\usepackage[hidelinks]{hyperref}
\usepackage{comment}
\usepackage{paralist}
\usepackage{pifont}

\usepackage{bbding}

\newcommand{\hide}[1]{}

\newcommand{\bit}{\begin{compactitem}}
\newcommand{\eit}{\end{compactitem}}
\newcommand{\ben}{\begin{compactenum}}
\newcommand{\een}{\end{compactenum}}

\usepackage{xcolor}

\usepackage{xcolor}

\usepackage{thmbox} 
\newtheorem[M, bodystyle=\normalfont\noindent]{definition}{Definition}
\newtheorem[M, bodystyle=\normalfont\noindent]{assumption}{Assumption}

\usepackage{thmtools}
\usepackage{thm-restate}
\usepackage{hyperref}

\usepackage{mathtools}

\ifCLASSINFOpdf
   \usepackage[pdftex]{graphicx}
\else
\fi
\makeatletter
\let\old@ps@headings\ps@headings
\let\old@ps@IEEEtitlepagestyle\ps@IEEEtitlepagestyle
\def\psccfooter#1{%
    \def\ps@headings{%
        \old@ps@headings%
        \def\@oddfoot{\strut\hfill#1\hfill\strut}%
        \def\@evenfoot{\strut\hfill#1\hfill\strut}%
    }%
    \def\ps@IEEEtitlepagestyle{%
        \old@ps@IEEEtitlepagestyle%
        \def\@oddfoot{\strut\hfill#1\hfill\strut}%
        \def\@evenfoot{\strut\hfill#1\hfill\strut}%
    }%
    \ps@headings%
}
\makeatother

\psccfooter{%
        \parbox{\textwidth}{\hrulefill \\ \small{24th Power Systems Computation Conference} \hfill \begin{minipage}{0.2\textwidth}\centering \vspace*{4pt} \includegraphics[scale=0.06]{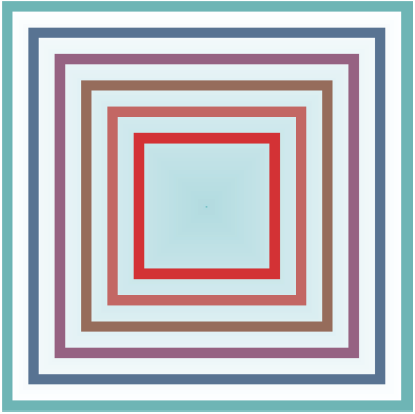}\\\small{PSCC 2026} \end{minipage} \hfill \small{Limassol, Cyprus --- June 8 -- 12, 2026}}%
}

\usepackage{tikz}
\usetikzlibrary{arrows.meta}

\begin{document}
\title{Cyber-Resilient System Identification for Power Grid through Bayesian  Integration
}

\author{
Shimiao Li$^1$,
    Guannan Qu$^2$, Bryan Hooi$^3$, Vyas Sekar$^2$, Soummya Kar$^2$, Larry Pileggi$^2$\\
$^1$EE Dept., University at Buffalo
shimiaol@buffalo.edu\\
$^2$ECE Dept., Carnegie Mellon University
\{gqu, vsekar, soummyak, pileggi\}@andrew.cmu.edu\\
$^3$National University of Singapore
bhooi@comp.nus.edu.sg\\
}

\maketitle

\begin{abstract}
	Power grids increasingly need real-time situational awareness under the ever-evolving cyberthreat landscape. Advances in snapshot-based system identification approaches have enabled accurately estimating states and topology from a snapshot of measurement data, under random bad data and topology errors. However, modern interactive, targeted false data can stay undetectable to these methods, and significantly compromise estimation accuracy. This work advances system identification that combines snapshot-based method with time-series model via Bayesian Integration, to advance cyber resiliency against both random and targeted false data. Using a distance-based time-series model, this work can leverage historical data of different distributions induced by changes in grid topology and other settings. The normal system behavior captured from historical data is integrated into system identification through a Bayesian treatment, to make solutions robust to targeted false data. We experiment on mixed random anomalies (bad data, topology error) and targeted false data injection attack (FDIA) to demonstrate our method's 1) cyber resilience: achieving over $70\%$ reduction in estimation error under FDIA; 2) anomalous data identification: being able to alarm and locate anomalous data; 3) almost linear scalability: achieving comparable speed with the snapshot-based baseline, both taking $<$1min per time tick on the large 2,383-bus system using a laptop CPU.
\end{abstract}
\thanksto{This manuscript has been submitted to PSCC2026.
}

\section{Introduction}\label{sec:intro}
For reliable operation in unpredictable environments, the electrical power grid requires real-time situational awareness (SA) that can accurately estimate the normal or abnormal state of the system, even with poor data qualities. The ever-evolving cyberthreat landscape makes it increasingly challenging. The grid today faces not only traditional errors which are sparse, random and independent (e.g., a random topology error of wrong line status), but also targeted cyber threats \cite{grid-security-Vyas}. In particular, advanced cyberattack models have been proved capable of manipulating demand values (e.g., MadIoT attack \cite{madiot}\cite{gridwarm}), causing data delays (e.g., jitter attack), and compromising data integrity in a stealthy way (e.g., false data injection attack, or FDIA\cite{fdia-review}). 

Significant advances have been made in developing robust methods for system identification under naturally-occurring random data anomalies. In this paper, system identification refers specifically to the estimation of network topology and AC system states (bus voltages), rather than other network parameters such as transmission line impedances. In modern control rooms, such an estimation is routinely performed from a single snapshot of measurements under the assumption of full observability \cite{SE_observability, grid_observability_analysis}\cite{SE_observability}\cite{grid_observability_analysis}. Traditional AC steady state estimation (SE) is based on weighted least squares (WLS) method \cite{WLS-SE} to solve bus voltages, followed by residual-based hypothesis tests to detect some random bad data. 
A variant \cite{TE-WLSE-BDI}  of this ran SE for a set of topologies and then applied residual tests to determine the optimal topology based on the one with the smallest residual. Generalized state estimation (GSE) methods \cite{GSE-DC-substation} \cite{GSE-AC-nonlinear} extended the WLS-based SE to estimate the topology jointly with bus voltages.
All of these traditional approaches are based on hypothesis tests to detect bad data, together with iteratively reweighting\cite{robustSE-reweight} and resolving to achieve accurate estimation results. 
To more efficiently detect and reject random errors, later works developed robust SE \cite{LAVSE-PMU-abur} and robust GSE\cite{TESE-GSE-PMUabur} using weighted least absolute value (WLAV) methods whose state solution is not affected by random errors and whose sparse residuals directly locate errors; however, they are based on unrealistic assumptions that the power system is fully observable with modern Phasor Measurement Unit (PMU) data. To adapt to the realistic data collection of both SCADA and PMU, circuit-theoretic estimators \cite{SUGAR-SE-Li}\cite{convexSE-LAV-Li}\cite{ckt-SE-TnD}\cite{ckt-GSE} were proposed: \cite{SUGAR-SE-Li} mapped SCADA and PMU sensors to linear circuits, giving linear formulations to include both; \cite{convexSE-LAV-Li}\cite{ckt-SE-TnD} extended it to robust SE which detects bad data while maintaining accurate states; and \cite{ckt-GSE} further extended it to robust GSE which also detects and rejects random topology errors.
Work in \cite{convexTESE-SDP-weng} also formulated a robust GSE considering hybrid SCADA and PMU data; however, the use of semidefinite programming (SDP) relaxation to convexify the problem results in a lack of scalability and efficiency on large-scale networks.

Unfortunately, the snapshot-based nature (i.e. relying on one most recent observation dataset without using the historical data series) leaves the above system identification approaches vulnerable to interactive or targeted errors that exploit physical laws, leading to very inaccurate state estimates when these threats occur.  Existing work has studied the extension of state estimation beyond a single snapshot, using a finite sequence of historical measurements. Notable examples include dynamic state estimation \cite{DSE} and moving-horizon state estimation \cite{MHSE}, which are typically built on transient dynamic models and/or with simple zero-centered disturbances modeled. The overall estimations are usually of higher complexity, and apply to short-term horizons (at second scale or shorter). Many methods benefit significantly from assuming high penetration of high-speed and high-quality PMUs, which is sometimes far from realistic. Moreover, their primary goal is usually to capture the fast dynamics of the system rather than to address data integrity issues for cyber-resilience.

How to retain accurate estimate under not only random data errors but also interactive targeted errors? This paper explores how cyber-resilient situational awareness can be advanced from a long-term steady-state horizon. We aim to extend the state-of-the-art steady-state system identification to incorporate long-term historical measurements, and account for two realistic constraints: 1) \textbf{hybrid data}: a mixture of steady-state data from SCADA and PMUs; and 2) \textbf{non-IID (independent and identically distributed) data}: historical observations collected under varying topologies and data conditions. While one might consider replacing model-based identification with a purely data-driven estimator trained on falsified data, this is extremely expensive and challenging, especially when accounting for topology changes and zero injection bus constraints. Thus, standalone learning-based estimators are beyond the scope of this study.

To achieve our goal, this work combines snapshot-based system identification with time-series processing of historical data. 
Starting from a state-of-the-art snapshot-based system
identification (which gives reliable estimations whenever no highly interactive or targeted false data occur), we leverage a distance-based time-series processing model to robustly learn normal system behavior from long-term non-IID historical state estimate data. The learned prior distribution that captures temporal context is then integrated into the system identification problem via Bayesian integration (specifically in the form of regularization), producing state estimates that remain accurate and resilient even under interactive or targeted data manipulations.

To evaluate the efficacy of proposed method, we experiment on IEEE standard power system transmission networks of different sizes (ranging from 30 buses to 3000+ buses) and test on a mixture of random anomalies (e.g., bad data or topology errors) and targeted false data injection attacks (FDIA). The results demonstrate the strength of the proposed method in 1) cyber resilience: significantly improving the accuracy of the estimation under targeted false data, achieving an over $70\%$ reduction in the estimation error under FDIA; 2) anomalous data identification: being able to alarm and locate different types of anomalous data; 3) almost linear scalability: achieving comparable speed with the snapshot-based baseline, both taking less than one minute per time tick on the large 2,383-bus system using a laptop CPU.

\section{Background: Anomalous grid data}

This Section provides the background of diverse data error types which form the motivation for developing cyber-resilient system identification.

At any given time, let $\bm{z}$ denote a snapshot of \textbf{steady-state} measurements collected across the system by both traditional SCADA and modern PMUs. A real-world snapshot $\bm{z}$ comprises both \emph{continuous} analog measurements and \emph{discrete} status data.
Each measurement $z_i$ can be modeled as
$z_i = h_i(\bm{v}) + \text{noise}$, where $h_i(\bm{v})$ is the physical relationship between the measurement and the system state $v$ (bus voltages), and measurement noises are typically modeled as zero-centered Gaussian $\mathcal{N}(0, \sigma^2)$.

In practice, measurements may suffer from different types of data quality issues:
\begin{enumerate}
    \item \textbf{Random data errors}: with uncorrelated, sparse data errors typically caused by communication faults and sensor noise. 
    For continuous data, these errors can take arbitrary magnitudes; 
    for discrete data, topology errors appear as random flips in line or breaker statuses.

    \item \textbf{Interactive or targeted false data}: correlated or coordinated manipulations in data values. 
    A notable example is the \emph{false data injection attack (FDIA)}~\cite{fdia-review}, where the attacker modifies measurements from $\bm{z}$ to 
    \begin{equation}
        \bm{z}' = h(\bm{v}') + \text{noise},
    \end{equation}
    misleading the operators into obtaining erroneous states $\bm{v}'$, while remaining undetectable to any snapshot-based system identification.
\end{enumerate}

\section{Spatial and temporal building blocks}\label{sec: building blocks}

This section presents the spatial (snapshot-based) and temporal building blocks separately. Building on our previous work, we extend these modules for broader and more realistic applications. We first discuss snapshot-based system identification and its extension to the commonly used bus-branch model via pseudo-switches. We then describe a distance-based temporal weighting approach for processing historical non-IID data. The next section will integrate these two building blocks through a Bayesian formulation.

\subsection{Robust snapshot-based system identification with bus-branch model compatibility}\label{sec: ckt-GSE}

When a snapshot of measurement $\bm{z}$ guarantees \textbf{(full) observability} of the system, the system state at this time moment can be uniquely determined. Prior work \cite{ckt-GSE} proposed ckt-GSE, a robust circuit-based generalized state estimator, to jointly estimate states and topology together, while identifying possible random errors in both the continuous data and status data. 

\begin{figure}[htbp]
     \centering
     \begin{subfigure}[b]{0.49\linewidth}
         \centering         \includegraphics[width=01\linewidth]{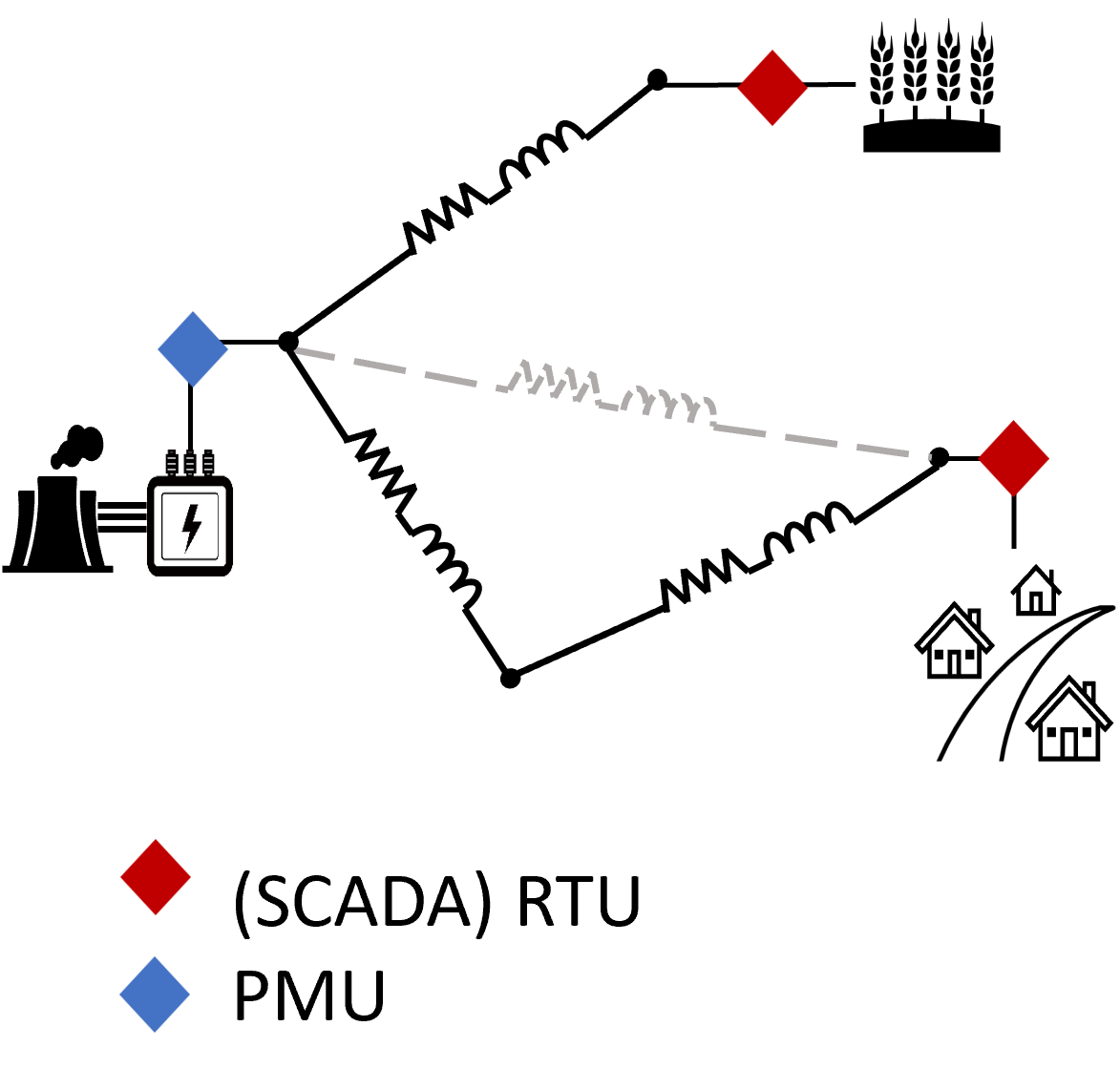}
         \caption{}
         \label{fig:case4}
     \end{subfigure}
     \hfill
     \begin{subfigure}[b]{0.49\linewidth}
         \centering         \includegraphics[width=01\linewidth]{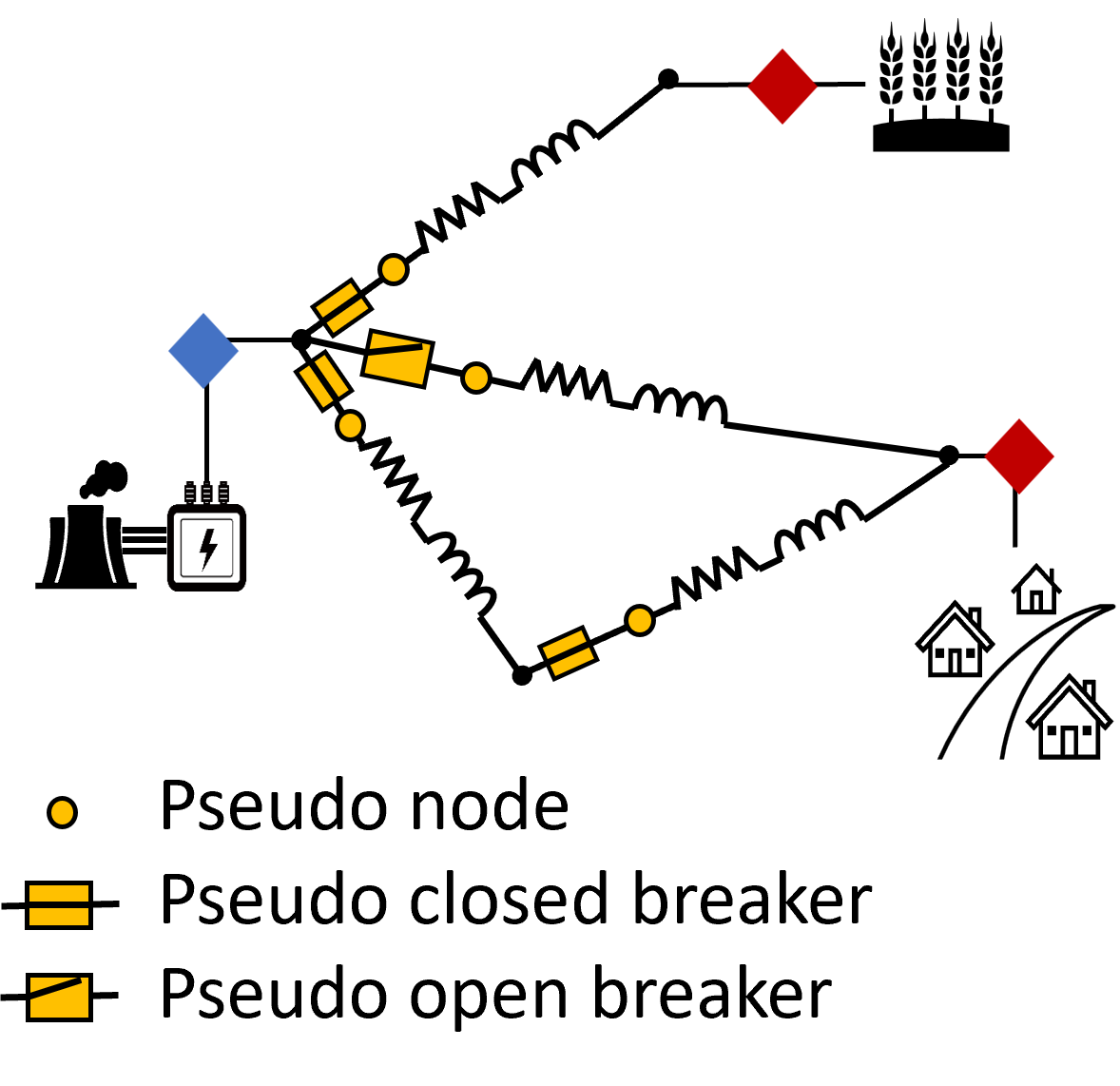}
         \caption{}
         \label{fig:case4 circuit model}
     \end{subfigure}
     \hfill
     \begin{subfigure}[b]{\linewidth}
         \centering         \includegraphics[width=01\linewidth]{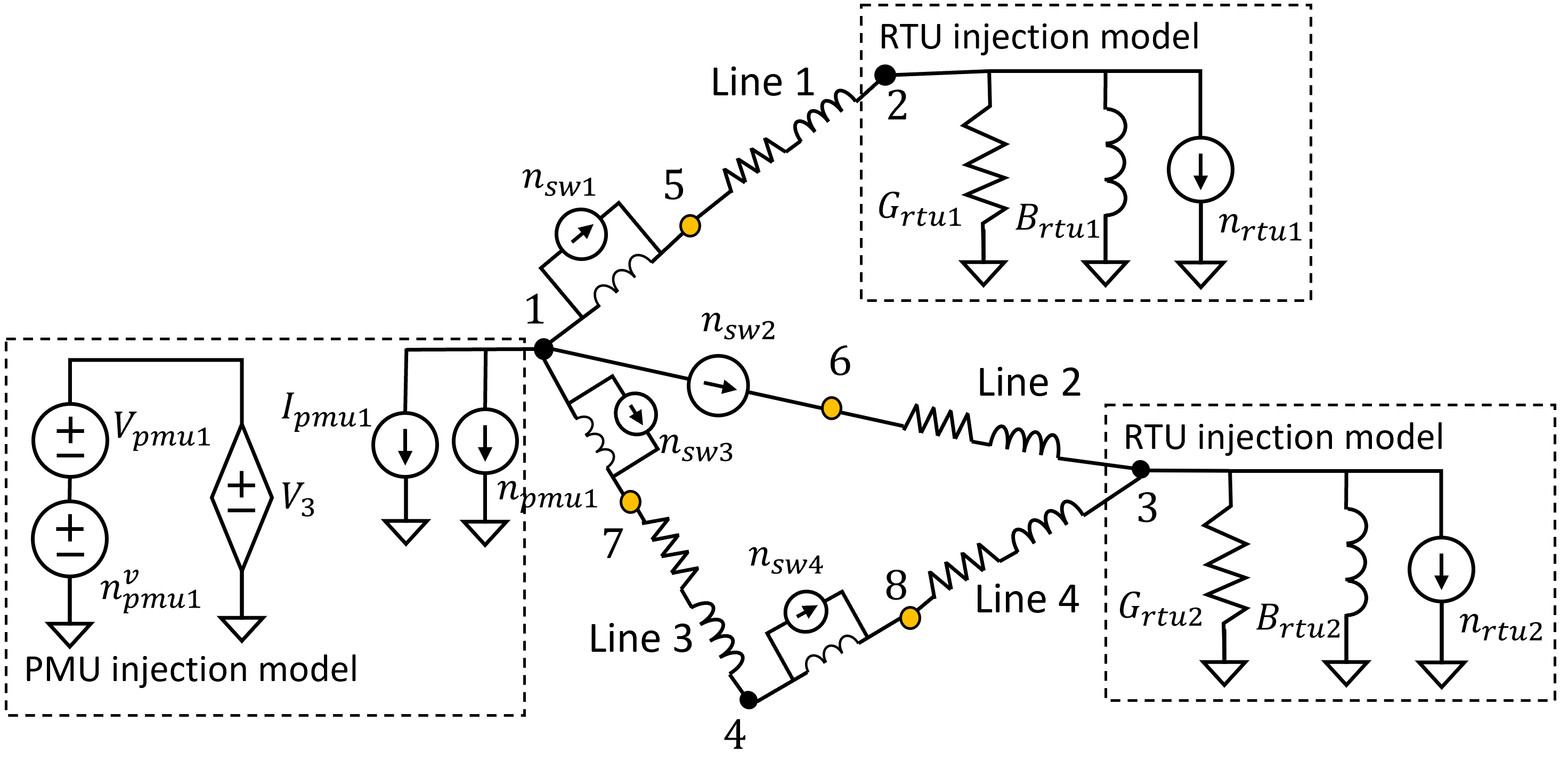}
         \caption{}
         \label{fig:case4 circuit model}
     \end{subfigure}
        \caption{Generalized state estimation on bus-branch model: (a) at any time $t$, given a power grid that is measured by SCADA and modern PMUs, the grey dashed transmission line is inactive; (b) if it is a bus-branch model, we add pseudo buses and pseudo circuit breakers, in order to estimate topology; the breaker statuses represent the associated line statuses, i.e., an active line connects to a closed breaker and an inactive line connects to an open breaker; (c) the measured elements are replaced with equivalent circuit models (PMU model, RTU model, and open/closed switch model, as developed in \cite{ckt-GSE}\cite{SUGAR-SE-Li}\cite{convexSE-LAV-Li}) to transform the power grid into an aggregated linear equivalent circuit on which the independent current sources $n_{pmu}, n_{rtu}, n_{sw}$ capture data errors and they are to be solved by
        minimization. The RTU model has $G_{rtu}=\frac{P_{rtu}}{|V|_{rtu}^2}; B_{rtu}=-\frac{Q_{rtu}}{|V|_{rtu}^2}$.}
        \label{fig: ckt-GSE}
\end{figure}

However, the original work\cite{ckt-GSE} was demonstrated on a node-breaker model that contains all bus sections, nodes, switches, and branches that are installed on the realistic grid. In practice, the majority of estimation, simulation, and optimization processes and studies on today's power grid rely on the bus-branch model. In this Section, we extend ckt-GSE to bus-branch model without loss of generality, as in Figure \ref{fig: ckt-GSE}. 

With each measurement modeled, ckt-GSE solves the weighted least absolute value (WLAV) estimation. In this problem, slack variables $n_{rtu}, n_{pmu},$ and $n_{sw}$ capture random sparse errors on the RTU, PMU, and line status data respectively; the state vector $v=[V^R_1,V^I_1,...,V^R_N,V^I_N]$ contains the estimation of real and imaginary bus voltages. 
\begin{subequations} \label{eq:GSE}
\begin{equation}
    \min_{\bm{v,n}}\sum_k w_k|n_{swk}|+\sum_i\alpha_i|n_{rtui}|\notag\\
    + \sum_j\beta_{j}|n_{pmuj}| 
    + \gamma_{j}|n_{pmuj}^{v}|
    \label{obj}
\end{equation}
$$
    {\text{s.t. (linear) KCL equations at all nodes: } \bm{A}\begin{bmatrix}
        \bm{v}\\\bm{n}
    \end{bmatrix}=\bm{b}}
$$
{\color{black}
Here, the constraint matrix 
$\bm{A}$ and vector 
$\bm{b}$ are constructed from the  measurements $\bm{z}$ and system parameters. Specifically, the measurement vector $\bm{z}$ consists of 1) phasor measurement unit (PMU) data, including voltage phasors $V_{pmu}^R + jV_{pmu}^I$ and current phasors $I_{pmu}^R + jI_{pmu}^I$; 2) SCADA data collected by remote terminal units (RTUs), including voltage magnitudes $|V|_{rtu}$ and real/reactive power measurements $P_{rtu}$ and $Q_{rtu}$; 
and 3) binary open/closed statuses of transmission lines, circuit breakers, and other switching devices. 
Taking Fig \ref{fig:case4 circuit model} as an example, the KCL (Kirchhoff's Current Law) equations at each location can be written as (\ref{eq: kcl example start})-(\ref{eq: kcl example end}), consisting of real (R) and imaginary (I) parts. Intermediate variables $I_{swk}$ denote current flowing through any switch $k$.\\
Node 1: (connects PMU and switches)
\begin{equation}
\begin{split}
    I_{pmu1}^R + n_{pmu1}^R+ I_{sw1}^R + I_{sw2}^R+ I_{sw3}^R =0\\
     I_{pmu1}^I + n_{pmu1}^I + I_{sw1}^I+ I_{sw2}^I+ I_{sw3}^I =0
     \label{eq: kcl example start}
\end{split}
\end{equation}
PMU voltage at node 1: (control circuit)
\begin{equation}
\begin{split}
    V_{pmu1}^R + n_{pmu1}^R - V_1^R = 0\\
    V_{pmu1}^I + n_{pmu1}^I - V_1^I = 0
\end{split}
\end{equation}
Node 2: (connects RTU and line)
\begin{equation}
\begin{split}
    G_{rtu1} V_2^R-B_{rtu1}V_2^I+n_{rtu1}^R +\\ G_{line1}(V_2^R-V_5^R) - B_{line1}(V_2^I-V_5^I) = 0\\
    G_{rtu1}V_2^I+B_{rtu1}V_2^R+n_{rtu1}^I +\\ G_{line1}(V_2^I-V_5^I) + B_{line1}(V_2^R-V_5^R)  = 0
\end{split}
\end{equation}
Open switches $sw_2$:  (control circuit):
\begin{equation}
    I_{sw2}^R = n_{sw2}^R, I_{sw2}^I = n_{sw2}^I
\end{equation}
Closed switches $sw_1, sw_3, sw_4$ (control circuit $swk=(i,j)$)
\begin{equation}
\begin{split}
    I_{swk}^R = \frac{1}{x_{sw}}(V_i^I-V_j^I) + n_{swk}^R,\\ 
    I_{swk}^I = -\frac{1}{x_{sw}}(V_i^R-V_j^R) + n_{swk}^I \label{eq: kcl example end}
\end{split}
\end{equation}
Constraints on other nodes can be written similarly.
}
\end{subequations}

\subsection{Distance-based time-series model for non-IID  data}\label{sec: dynwatch}
Prior work DynWatch \cite{dynwatch} proposed a distance-based method to learn and characterize normal system behavior from historical time-series data under topology changes. This section generalizes that framework to account for broader variations in system behavior beyond topology changes.

At any time $t$, let us model the temporal behavior of any system state (and/or power flow) $x^t$ as a Gaussian distribution $x^t\sim\mathcal{N}^t =\mathcal{N} (\mu,\delta^2)$. Given $N$ historical records $x^1,...,x^N$ of $x$ values, it is intuitively easy to characterize this distribution using simple statistics, like calculating median or mean for $\mu$, and interquartile range (IQR) or variance for $\delta^2$. This is based on assuming the historical data are IID, i.e., independent and identically distributed.
    
However, historical grid data in practice are typically not IID, since the changes in grid topology and other configuration can give rise to different distributions.
When characterizing a distribution $\mathcal{N}^t$ for time $t$, it is desirable to give more trust to data from the same distribution or from the nearest distributions if not enough data are available, while neglecting those far away from $\mathcal{N}^t$. 

Intuitively, we characterize the distribution from weighted historical data, assigning weights $\bm{w}=[w^1, ..., w^N]^\intercal$ to historical data $x^1,..., x^N$, where we assign higher weights to data from nearer distributions and zero weights to those from far-away distributions. This can be achieved with a \textbf{distance-based time-series processing}.
Suppose data sample $k$ comes from distribution $\mathcal{N}^k$, and let $d^k$ denote the distance between distribution $\mathcal{N}^k$ and the target distribution $\mathcal{N}^t$, i.e. $d^k=Distance(\mathcal{N}^k, \mathcal{N}^t)$. This distance measure can be defined from graph edit distance, power system sensitivities factors \cite{dynwatch}, projections \cite{dynwatch-extended-Sangwoo}, or graph embedding techniques, .etc. 

With distance calculated, the use of data sample $k$ will introduce an empirical bias of up to $w_kd_k$, to the total statistical error. If we simply assign weights by minimal bias, we tend to use only the data from the nearest distribution, which might lead to limited data and overfitting, introducing variance (by adding L2 regularization $||\bm{w}||_2^2$). Therefore, the optimal assignment of sparse weights can be made by a bias-variance trade-off optimization:
\begin{equation}
\min_{w}{\underbrace{\sum_{k=1}^N w_k d_k}_\text{bias} + \underbrace{ \alpha||\bm{w}||_2^2}_\text{variance}}
\text{, s.t. } \underbrace{w_k\geq 0,  \forall k; \sum_{k}w_k = 1, }_\text{non-negative and sum to 1}
\label{fobj: bias-var trade-off}
\end{equation}
Or equivalently $\min_{\bm{w}} \sum_k d_k|w_k|+\alpha||\bm{w}||_2^2$ s.t. $\sum_k|w_k|=1$ where the weight $|w_k|$ values are encouraged to be sparse by minimizing the L1-norm $\sum_k d_k|w_k|$.

Upon weight assignment, we can characterize the distribution $\mathcal{N}^t=\mathcal{N}(\mu,\delta^2)$ using weighted historical samples. The weighted median of these historical samples defines the distribution center $\mu$ and the weighted IQR defines the dispersion $\delta$. The use of robust statistics makes the distribution characterization robust to anomalous samples or outliers in the historical data.


\section{Hybrid Design with Bayesian Integration}

Building upon the spatial and temporal building blocks presented in Section \ref{sec: building blocks}, this paper uniquely combines the two within a Bayesian framework to advance the cyber resilience of system identification under both random and targeted false data. Without such integration, snapshot-based system identification methods remain vulnerable to targeted false data, limiting their effectiveness in adversarial environments.

\subsection{Bayesian Integration}
\begin{figure}[h]
    \centering
\includegraphics[width=0.8\linewidth]{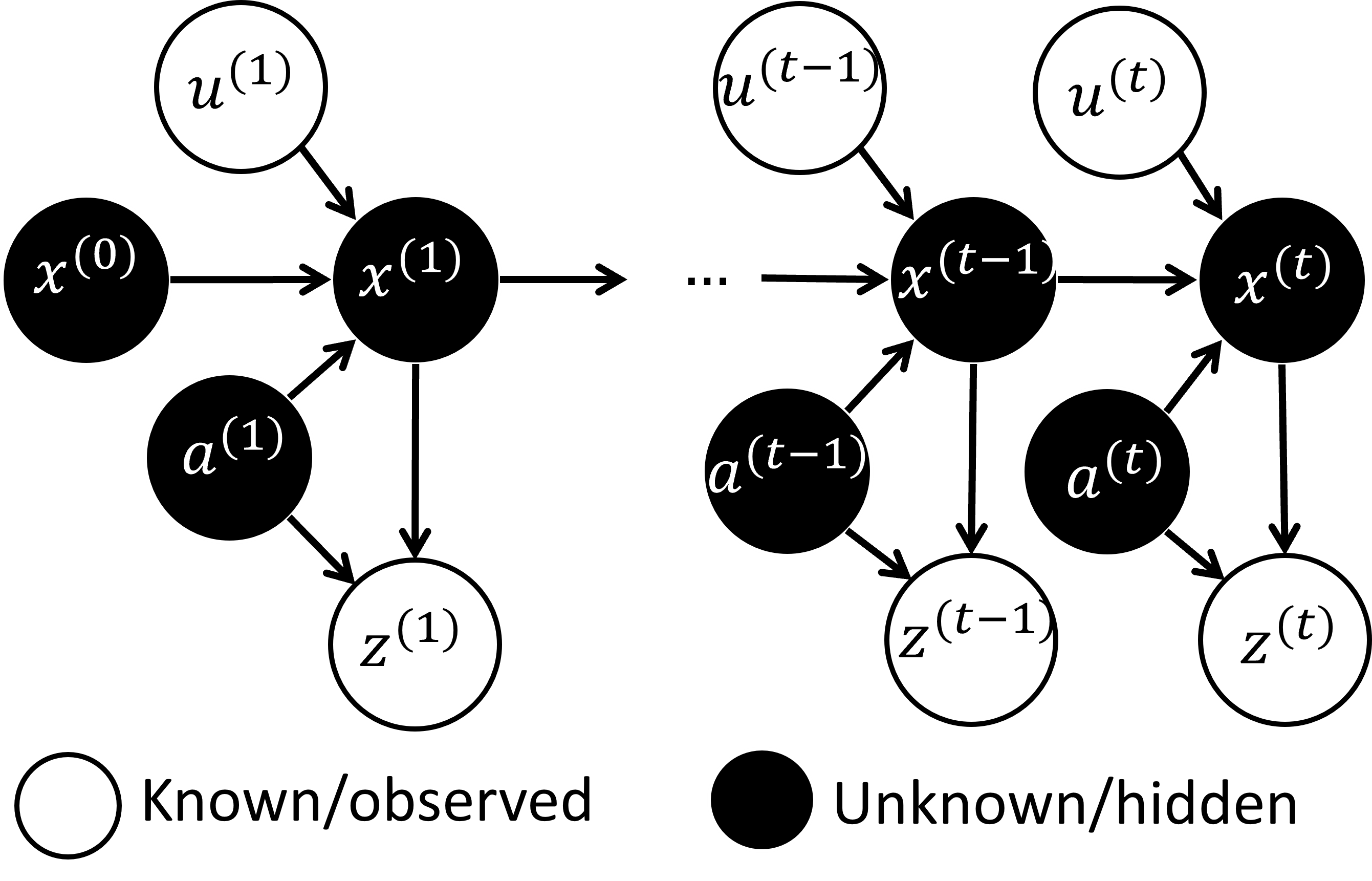}
    \caption{Bayesian Network: The system starts from an initial state (vector) $\bm{x}^{(0)}$ which includes the AC bus voltages $\bm{v^}{(0)}$ and network topology $G^{(0)}$. At any time $t$, control action $\bm{u^}{(t)}$ and stochastic process drive the system to change from $\bm{x}^{(t-1)}$ to new state $\bm{x}^{(t)}$; sensors collect real-time data (vector) $\bm{z^}{(t)}$; and anomalies $\bm{a^}{(t)}$ can affect the system's state and/or its data. Gaining situation awareness is to infer the unknown variables $\bm{x,a}$ given the known or observed variables $\bm{z,u}$.
    }
    \label{fig: pgm}
\end{figure}

The temporal grid evolution, along with the system data and anomalies, can be represented using a Bayesian network in
Figure \ref{fig: pgm}. Specifically, given historical measurements  $\bm{z^}{(1)},...,\bm{z^}{(t)}$, a trace of (known) input control signals $\bm{u^}{(0)},...,\bm{u^}{(t)}$, we can infer the AC states $\bm{v^}{(1)},...,\bm{v^}{(t)}$ (which correspond to bus voltages), network topology $G^{(1)},...,G^{(t)}$, and anomaly indicators $\bm{a^}{(1)},...,\bm{a^}{(t)}$.

For simplicity, let 
  $\bm{x}^{(t)}=[\bm{v^}{(t)},G^{(t)}]$  $\forall t$ include both AC states of bus voltages $\bm{v^}{(t)}$ and network topology $G^{(t)}$, since they can be jointly estimated.
 and let  $\widetilde{Z},\widetilde{U},\widetilde{X},\widetilde{A}$ denote the time-series of variables $\widetilde{Z}=[\bm{z^}{(1)},...,\bm{z^}{(t)}], 
 \widetilde{U}=[\bm{u^}{(1)},...,\bm{u^}{(t)}], 
\widetilde{X}=[\bm{x}^{(1)},...,\bm{x}^{(t)}], 
  \widetilde{A}=[\bm{a^}{(1)},...,\bm{a^}{(t)}]$, we estimate $\widetilde{V},\widetilde{G},\widetilde{A}$ via an Maximum A Posteriori (MAP) estimation which can be further decomposed after applying Bayesian factorization to $p(\widetilde{X},\widetilde{A},\widetilde{Z},\widetilde{U})$:
\begin{align}
&\widetilde{X*}, \widetilde{A*}=\arg\max p(\widetilde{X},\widetilde{A}|\widetilde{Z},\widetilde{U})\\
&=\arg\max \log p(\widetilde{X},\widetilde{A},\widetilde{Z},\widetilde{U})\notag\\
&=\arg\max \log\Biggl(p(\bm{v^}{(0)})\cdot\notag\\& \prod_{
\tau=1}^{t} P(\bm{a^}{(t)})P(\bm{u^}{(t)})
p(\bm{x}^{(t)}|\bm{x}^{(t-1)},\bm{u^}{(t)}, \bm{a^}{(t)})p(\bm{z^}{(t)}|\bm{x}^{(t)},\bm{a^}{(t)})\Biggr) \notag\\
&=\arg\max \log p(\bm{v^}{(0)}) +\sum_{\tau=1}^{t} 
 \log p(\bm{a^}{(t)}) + \sum_{\tau=1}^{t} \log p(\bm{u^}{(t)}) \notag\\
 &+\sum_{\tau=1}^{t} \log p(\bm{z^}{(t)}|\bm{x}^{(t)},\bm{a^}{(t)}) + \sum_{\tau=1}^{t} \log p(\bm{x}^{(t)}|\bm{x}^{(t-1)},\bm{u^}{(t)},\bm{a^}{(t)}) \notag
\end{align}

Then we make assumptions on the priors that  $\forall t$, prior $p(\bm{v^}{(0)})$ is a continuous uniform distribution, because, intuitively, for any episode of state transition, the power grid can start at any random initial state.

Let us further narrow down our discussion of actions $\bm{u^}{(t)}$ to only include topology change actions.
For $\forall t$, prior $P(\bm{a^}{(t)})$ and $P(\bm{u^}{(t)})$ can be modeled as discrete uniform distributions. For $P(\bm{a^}{(t)})$, the uniform distribution is assumed because this paper focuses on point outliers and treats the anomaly $\bm{a^}{(t)}$ as independent of other time moments, and we assume no prior knowledge on the anomaly occurrence. Similarly, for $P(\bm{u^}{(t)})$, we assume no prior knowledge or preference on actions, thus assuming a uniform distribution to allow for any random topology change to occur at any time.

These assumptions reduce the estimation to the follows:
\begin{align}
&\bm{x}^{(1)*},...,\bm{x}^{(t)*},\bm{a^}{(1)*},...,\bm{a^}{(t)*}
=\arg\max \label{prob: bayesian estimation 1 to t}\\&
\underbrace{\sum_{\tau=1}^t \log p(\bm{z^}{(\tau)}|\bm{x}^{(\tau)},\bm{a^}{(\tau)}) }_\text{spatial}
+ \underbrace{\sum_{\tau=1}^t \log p(\bm{x}^{(\tau)}|\bm{x}^{(\tau-1)},\bm{u^}{(\tau)},\bm{a^}{(\tau)})}_\text{temporal}\notag
\end{align}
where the first term is a system identification problem which relies on the spatial relationship that depicts how the data $\bm{z^}{(t)}$ at time $t$ can be generated, whereas the second is a temporal state transition describing how topology control actions and anomalies cause grid state to change.   

The first term $\sum_t p(\bm{z^}{(\tau)}|\bm{x}^{(\tau)},\bm{a^}{(\tau)})$ can be easily written in system identification. Typically for $\forall t$, the measurement model for $\bm{z^}{(t)}$ can be characterized by $\bm{z^}{(t)}=h(\bm{x}^{(t)})+error/noise$ where the noise or error distribution determines $p(\bm{z^}{(t)}|\bm{x}^{(t)},\bm{a^}{(t)})$. Using the ckt-GSE method in Section \ref{sec: augmented ckt-GSE}, data error $n$ is assumed to be sparsely distributed (intuitively following a Laplace distribution).

Unfortunately, the temporal term $\sum_\tau p(\bm{x}^{(\tau)}|\bm{x}^{(\tau-1)},\bm{u^}{(\tau)},\bm{a^}{(\tau)})$ is not directly available. Power flow simulation can give an accurate estimate, but only when $\bm{x}^{(t-1)}$ is accurate, and it is a nonlinear programming problem that reduces time efficiency. Some existing works have treated the temporal steady-state transition as a simple 1st order Markov Chain $\bm{x}^{(t)}=\bm{x}^{(t-1)}+\epsilon$ with $\epsilon\sim N(0,\bm{\delta}^2)$, but these cannot explicitly account for changes in grid configuration (like the network topology). 

This work, instead, adopts a fast distance-based time-series model (in Section \ref{sec: dynwatch}) that predicts a distribution $q_{x(t)}$ from the  historical AC state and topology estimates to approximate the true distribution $p(\bm{x}^{(t)}|\bm{x}^{(t-1)},\bm{u^}{(t)},\bm{a^}{(t)})$ while accounting from grid changes. I.e.,
\begin{equation}
q_{x(t)}=q(\bm{x}^{(t)}|\bm{x}^{(t-1)},...,\bm{x}^{(1)},\bm{a^}{(t)},...,\bm{a^}{(1)})
\end{equation}

Thereby, the Bayesian inference  problem can be solved by uniquely combining the spatial and temporal building blocks that has been discussed in Section \ref{sec: building blocks}, which is not explored in prior works that treated them separately. This  remains computationally feasible since both components rely on lightweight convex optimizations with sparsity-promoting regularization, making the overall procedure suitable for real-time or near-real-time implementation. Motivated by this, we start by running snapshot-based system identification (which provides reliable estimates whenever no targeted false data occur), and then proceed with the following steps:
\begin{enumerate}
    \item \textbf{Time-series prediction} to output distribution $q_{x(t)}$ as an approximation of $p(\bm{x}^{(t)}|\bm{x}^{(t-1)},\bm{u^}{(t)},\bm{a^}{(t)})$ for each $t$: the prediction is conditioned on the most updated $\bm{x}^{(t)},G^{(t)}$ for $\forall t$ from system identification. This paper uses distance-based time-series processing (DynWatch) as the base model due to its ability to consider non-IID time-series data. 
    \item \textbf{Augmented system identification}: given $q_{x(t)}$, the estimation problem in (\ref{prob: bayesian estimation 1 to t}) reduces to:
    \begin{align}
&\bm{x}^{(1)*},...,\bm{x}^{(t)*},\bm{a^}{(1)*},...,\bm{a^}{(t)*}\\
=&\arg\max 
\underbrace{\sum_{\tau=1}^t \log p(\bm{z^}{(\tau)}|\bm{x}^{(\tau)},\bm{a^}{(\tau)}) }_\text{spatial (system identification)}
+ \underbrace{\sum_{\tau=1}^t log q_{x(\tau)}}_\text{temporal (prior knowledge)}\notag
\end{align}
or equivalently, for $\forall t$,
\begin{equation}
    \bm{x}^{(t)*},\bm{a^}{(t)*}
=\arg\max 
\underbrace{\log p(\bm{z^}{(t)}|\bm{x}^{(t)},\bm{a^}{(t)}) }_\text{system identification}
+ \underbrace{log q_{x(t)}}_\text{temporal prior}
\label{prob: augmented SI, probabilistic concept}
\end{equation}
This can be considered as a system identification augmented by prior knowledge from temporal pattern. This paper adopts ckt-GSE as the base model to be augmented, due to its advantages over other baseline estimators. The augmentation is explained in Section \ref{sec: augmented ckt-GSE}. 
\end{enumerate}

\subsection{ckt-GSE augmented by prior knowledge}\label{sec: augmented ckt-GSE}
As illustrated in (\ref{prob: augmented SI, probabilistic concept}), the first part of the estimation problem:$
    \bm{x}^{(t)*},\bm{a^}{(t)*}=\arg\max 
{\log p(\bm{z^}{(t)}|\bm{x}^{(t)},\bm{a^}{(t)}) }$
can be seen as a snapshot-based system identification problem for time $t$. Using variables $n$ to capture anomalies at each measurements, ckt-GSE which solves $\bm{v^}{(t)*}, \bm{n}^{(t)*}=arg\min ||W\bm{n}^{(t)}||_1 \text{ subject to } \bm{A^}{(t)}\begin{bmatrix}
    \bm{v^}{(t)}\\\bm{n}^{(t)}
\end{bmatrix}=\bm{b}^{(t)}$ (as introduced in Section \ref{sec: ckt-GSE}) can be seen as a deterministic relaxation of this probabilistic model, intuitively corresponding to the assumption that errors $\bm{n}$ follow independent Laplace distributions.  

Now we consider the augmented system identification in (\ref{prob: augmented SI, probabilistic concept}) which introduces temporal prior knowledge to advance its robustness. Let us make the following assumption made on the prior knowledge of AC state vector $\bm{v^}{(t)}$:

\begin{assumption}[state prior]
For a bus/node $i$ at time $t$, its AC state $v_i^{(t)}=[V_i^{(t),R}, V_i^{(t),I}]$ is expected to take the value around $\mu_i^{(t)}=[\mu_i^{(t),R}, \mu_i^{(t),I}]$ with uncertainty (standard deviation) $\delta_i^{(t)}$. Consider the prior distribution as Gaussian, we have 
\begin{equation}
    \bm{v^}{i,t}=\begin{bmatrix}
V_i^{(t),R}\\ V_i^{(t),I}
\end{bmatrix}\sim N(\begin{bmatrix}
\mu_i^{(t),R}\\ \mu_i^{(t),R}
\end{bmatrix}, \begin{bmatrix}
\delta_i^{(t)2},0\\
0,\delta_i^{(t)2}\\
\end{bmatrix})
\end{equation}
or, equivalently, the state vector $\bm{v^}{(t)}$ at time $t$ has a Gaussian prior distribution $q_{v(t)}$ such that
\begin{equation}
   \bm{v^}{(t)}\sim N(\bm{\mu^}{(t)}, \Delta^{(t)}) 
\end{equation}
with $\Delta^{(t)}$ being a diagonal matrix specifying the covariance matrix of the prior.
\end{assumption}

Taking this state prior $q_{v(t)}$, the augmented system identification can be defined as below:
\begin{definition}[augmented ckt-GSE with state prior] 
\label{def: augmented ckt-GSE}
At time $t$, given  $q_{v(t)}$ which is a prior knowledge of AC states $\bm{v^}{(t)}$, the augmented ckt-GSE runs a convex constrained optimization problem ($\beta=0.1$ used in this work):
\begin{align}   &\min_{v,n}\underbrace{||W\bm{n}^{(t)}||_1}_\text{ckt-GSE objective} + \underbrace{\beta\cdot (\bm{v^}{(t)}-\bm{\mu^}{(t)})^\intercal  (\Delta^{(t)})^{-1}(\bm{v^}{(t)}-\bm{\mu^}{(t)})}_\text{state prior}\notag\\
    &\text{ subject to } \bm{A}^{(t)}\begin{bmatrix}
    \bm{v^}{(t)}\\\bm{n}^{(t)}
\end{bmatrix}=\bm{b}^{(t)}
\end{align}
\end{definition}

\section{Results}\label{sec:experiments}
\label{sec:Results}
To validate the efficacy of the proposed method, we design experiments to answer the following questions:
\begin{enumerate}
    \item \textbf{Anomaly identification:} Is the proposed method able to identify the random and targeted false data?
    \item \textbf{Cyber resilience:} Is the state and topology estimation  robust to random and targeted data errors? (i.e., can accuracy be maintained?) 
    \item \textbf{Speed and scalability:} What is the worktime on different sized networks?
\end{enumerate}
\noindent 

We test and compare two methods:
\begin{itemize}
    \item \textbf{(Baseline) ckt-GSE}: which solves a snapshot based system identification for each time moment.
    \item \textbf{(Proposed) multi-period ckt-GSE}: which first solves ckt-GSE, followed by time-series processing and augmented ckt-GSE (as proposed in Section \ref{sec: augmented ckt-GSE}.
\end{itemize}
on the following different sized systems:
\begin{enumerate}
\item \textbf{Case30}\cite{case30}: 30-bus test case which represents a simple approximation of the American Electric Power system.
    \item \textbf{Case1354pegase}\cite{case1354pegase}: a 1354-bus network which accurately represents the size and complexity of part of the European high voltage transmission network.
   \item \textbf{Case2383wp}\cite{case2383wp}: part of the 7500+ bus 
   European power system which represents the Polish 400, 220 and 110 kV networks during winter 1999-2000 peak conditions.
   \item \textbf{Case2869pegase} \cite{case2869pegase_1}\cite{case2869pegase_2}: a 2869-bus network representing the size and complexity of part of the
   European high voltage transmission network.
\end{enumerate}

All our experiments are run on a laptop computer with 11th Gen Intel(R) Core(TM) i7-1185G7 @ 3.00GHz  1.80 GHz processor and 32 GB RAM. 

\textbf{Anomaly settings:} We tested on 3 types of anomalous data:
\begin{enumerate}
    \item (Random) bad data: we add large deviation (1 per unit) to continuous data values on 2 randomly selected locations.
    \item (Random) topology error: we modify status data at 2 random lines to make the operator have a wrong graph structure (e.g. a transmission line switch is actually closed but now reported as open; or actually open but now reported as closed).
    \item Targeted false data using false data injection attack (FDIA) model: we assume attackers are capable of compromising any data location, and modifying data values to mislead the operators into thinking the load demand decreased by 20\%.
\end{enumerate}

\textbf{Time-varying topology:} The time-series data in this paper is created with both known topology changes, both regular and irregular. We simulate regular topology changes occurring every 60 time ticks. In each instance, a previously open transmission line becomes closed, and meanwhile a new (random) line becomes open. We also simulate sudden line opening at random time moments and only lasting for 1 time tick. These irregular topology changes represent real-world line outages, and we assume that the status data can report these changes so that operators can quickly reconnect the line to fix the problem. (Any line outage not accurately reported by the status data is considered a topology error as described in anomalous data.)

\textbf{Time-series data settings:} We simulate (steady-state) time-series data (600 time ticks for case30, and 300 for cases), with time-varying topology as described above. We assume PMUs are installed on every generation bus measuring $V, I$ phasors; traditional SCADA RTUs are installed on every load bus measuring $|V|, P, Q$; RTU line meters are installed on  selected transmission lines measuring line power flow $P_{line}, Q_{line}$ and $|V|$ at one end of the line. The measurement data are generated by simulating power flow and adding random Gaussian noise to the power flow results. The random and targeted anomalous data (described above) are inserted on randomly selected moments.

\subsection{Anomaly identification}
\begin{figure}[h]
	\centering
\includegraphics[width=0.99\linewidth]{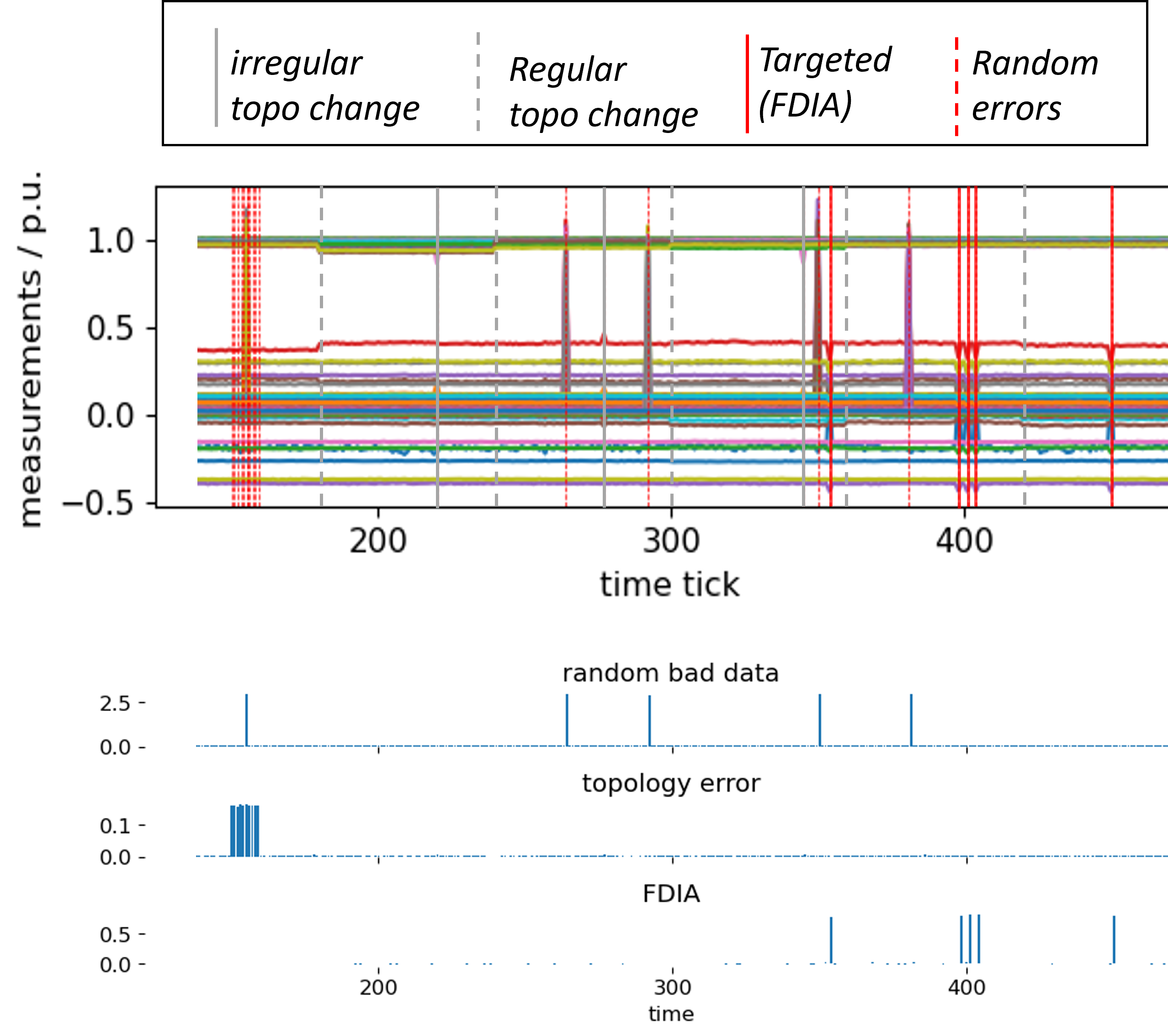}
     \caption{Anomaly detection on 30-bus system: bars plot the magnitude of $\bm{n}$. Top: Time-series measurements on case30: (a subset of the 600 time ticks). Bottom: random bad data are detectable by $|\bm{n}_{pmu}|,|\bm{n}_{rtu}|$ from ckt-GSE; topology errors are detectable by $|\bm{n}_{sw}|$ from ckt-GSE; targeted false data injected by FDIA are detectable by $|\bm{n}|$ from augmented ckt-GSE. These different anomalous data are distinguishable from a combined used of ckt-GSE and augmented ckt-GSE in this work. See Figure \ref{fig: 1354pegase AD} for detection on 1354-bus system.}
     \label{fig: toy AD}
     \end{figure}

Results in Figure \ref{fig: toy AD} shows the detection of anomalies and Figure \ref{fig: result, anomaly identification} further shows the localization of data anomalies, on a 30-bus system. Results on a larger 1354-bus system are shown in Figure \ref{fig: 1354pegase AD} and Figure \ref{fig: 1354pegase, anomaly localization} in the Appendix.

 By comparison, the baseline ckt-GSE, as a snapshot-based system identification, is limited to locating random bad data and topology errors; whereas the augmented ckt-GSE leverages prior knowledge from historical states to further enable identifying FDIA cyberattacks and pinpointing the bus locations whose state estimates are most affected.
     
 \begin{figure}[htbp]
\includegraphics[width=\linewidth]{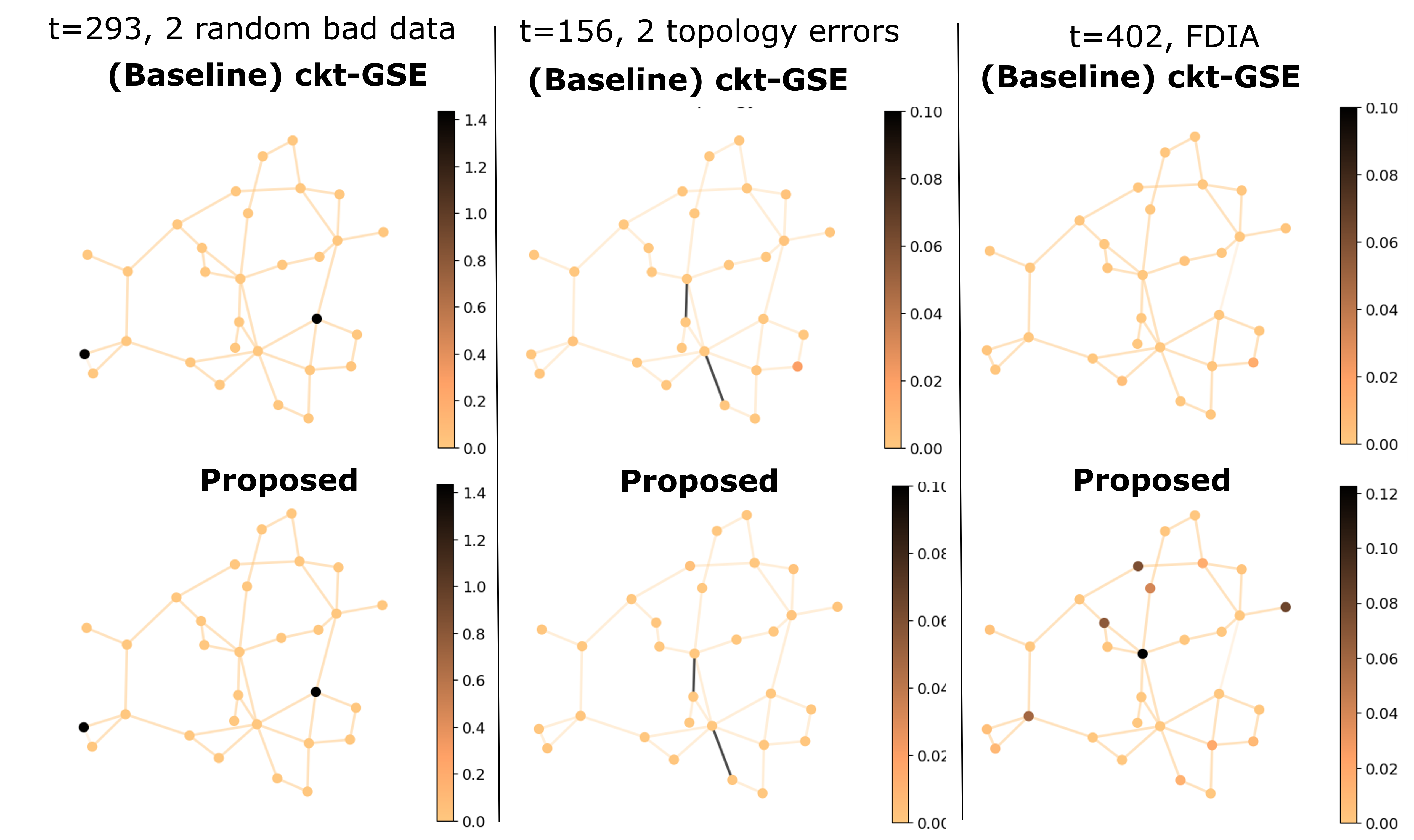}
        \caption{Anomaly localization on 30-bus system: baseline ckt-GSE locates the random bad data and topology errors; but fails to recognize FDIA. The proposed method identifies all of these threats. See Figure \ref{fig: 1354pegase AD} for result on 1354-bus system.}
        \label{fig: result, anomaly identification}
        \hfill
\end{figure}

\subsection{Cyber resilience / robustness}
We evaluate accuracy of state solution using root mean squared error $RMSE=\sqrt{\sum_{\text{bus } i}(v_{pred,i}-v_{true,i})^2}$. 
Whenever a false data injection attack happens, the solutions of ckt-GSE (when used alone) are significantly perturbed by falsified data; whereas having an augmented ckt-GSE serves to protect accuracy against FDIA. Figure \ref{fig: result, accuracy} illustrates the result.

\begin{figure}[htbp]
     \begin{subfigure}[h]{0.49\linewidth}
         \centering         
\includegraphics[width=\textwidth]{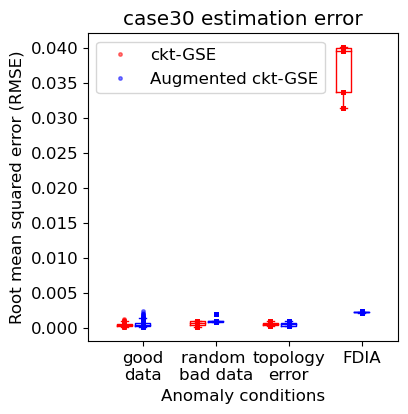}   \end{subfigure}
     \hfill
     \begin{subfigure}[h]{0.49\linewidth}
         \centering        
         \includegraphics[width=\textwidth]{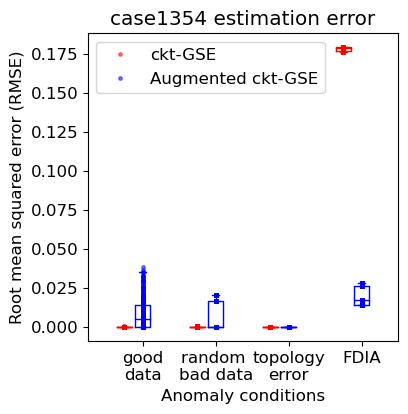}
     \end{subfigure}\\
     \begin{subfigure}[h]{0.49\linewidth}
         \centering       
\includegraphics[width=\textwidth]{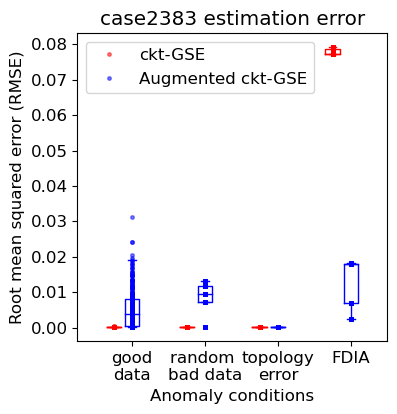}         
     \end{subfigure}
     \hfill
     \begin{subfigure}[h]{0.49\linewidth}
         \centering       
         \includegraphics[width=\textwidth]{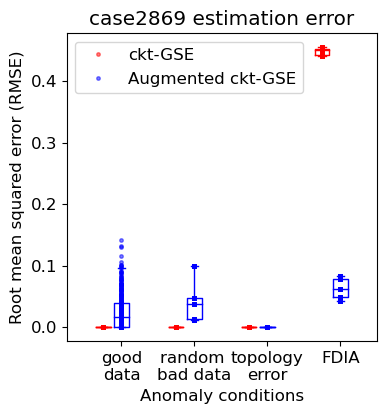}      
     \end{subfigure}\\
     \caption{Estimation robustness: Under FDIA, the augmented ckt-GSE significantly reduces estimation error, mitigating the adversary’s intended impact. But this comes with slightly lower accuracy under normal or random error conditions. The combined use of two estimators can address this: when large $\bm{n}$ from augmented ckt-GSE indicates targeted high penetration of false data, augmented ckt-GSE should be trusted; otherwise, baseline ckt-GSE provides the most accurate estimate.}
     \label{fig: result, accuracy}
     \end{figure}

\subsection{Speed and scalability}
The method needs to be time-efficient on large-scale networks to be applicable in real-world control rooms. Here we evaluate the speed and scalability of proposed work by evaluating its 3 essential steps: ckt-GSE, augmented ckt-GSE and (DynWatch) time-series processing. 

In our experiment, both ckt-GSE and augmented ckt-GSE are solved using the circuit-based linear programming (LP) solver developed in \cite{ckt-GSE}. Prior work \cite{ckt-GSE} has already shown that 
ckt-GSE which used our circuit-based LP solver is significantly faster than standard interior-point (IP) solver in python CVXOPT toolbox and Simplex method in SciPy which solves min-max model, especially on large scale cases. Prior work in \cite{dynwatch} has also shown that the distance calculation and time-series processing in DynWatch scales approximately linearly when used for anomaly detection. 

\begin{figure}[ht]
	\centering
	\includegraphics[width=0.99\linewidth]{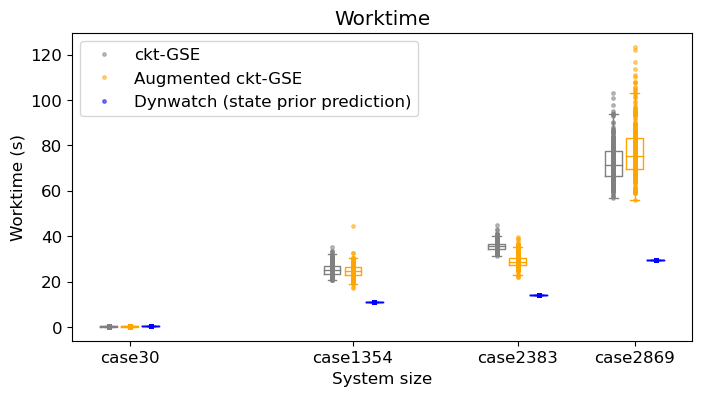}
	\caption[]{\color{black} Speed and scalability: run time scales well with system size. System size $=$ num of nodes $+$ num of edges in the network. Runtime of DynWatch is the average time needed per time tick to predict state prior for all buses.
     } 
	\label{fig:scalability}
\end{figure}

Figure \ref{fig:scalability} shows the worktime of the 3 essential steps on different sized networks. Results shows that these essential steps scale well. Meanwhile, the speed of augmented ckt-GSE and ckt-GSE are approximately the same, meaning that Bayesian integration advances system identification without introducing significantly computation burdens. 

\section{Conclusion}
\label{sec:Conclusion}
This work proposes a multi-period steady-state system identification that combines snapshot-based system identification with time-series processing to advance cyber resilience (or robustness) against both random and interactive / targeted false data. The use of distance-based time-series processing enables leveraging historical steady-state measurement data that come from different distributions (induced by changes in topology and other configurations). As a result, the proposed method is able to maintain accurate state estimates while identifying both random anomalies (bad data, topology error, etc) and targeted false data injections. Experiment results demonstrated the method’s strength in 1) cyber resilience: achieving over $70\%$ reduction in error under FDIA; 2) detection capability: being able to alarm on and locate different types of anomalous data; 3) almost linear scalability: achieving comparable speed with the snapshot-based baseline, both taking less than one minute per time tick on the large 2,383-bus system using a laptop CPU.


\bibliographystyle{IEEEtran}
\bibliography{BayesGSE}

\appendix
\begin{figure}[htbp]
	\centering
\includegraphics[width=0.99\linewidth]{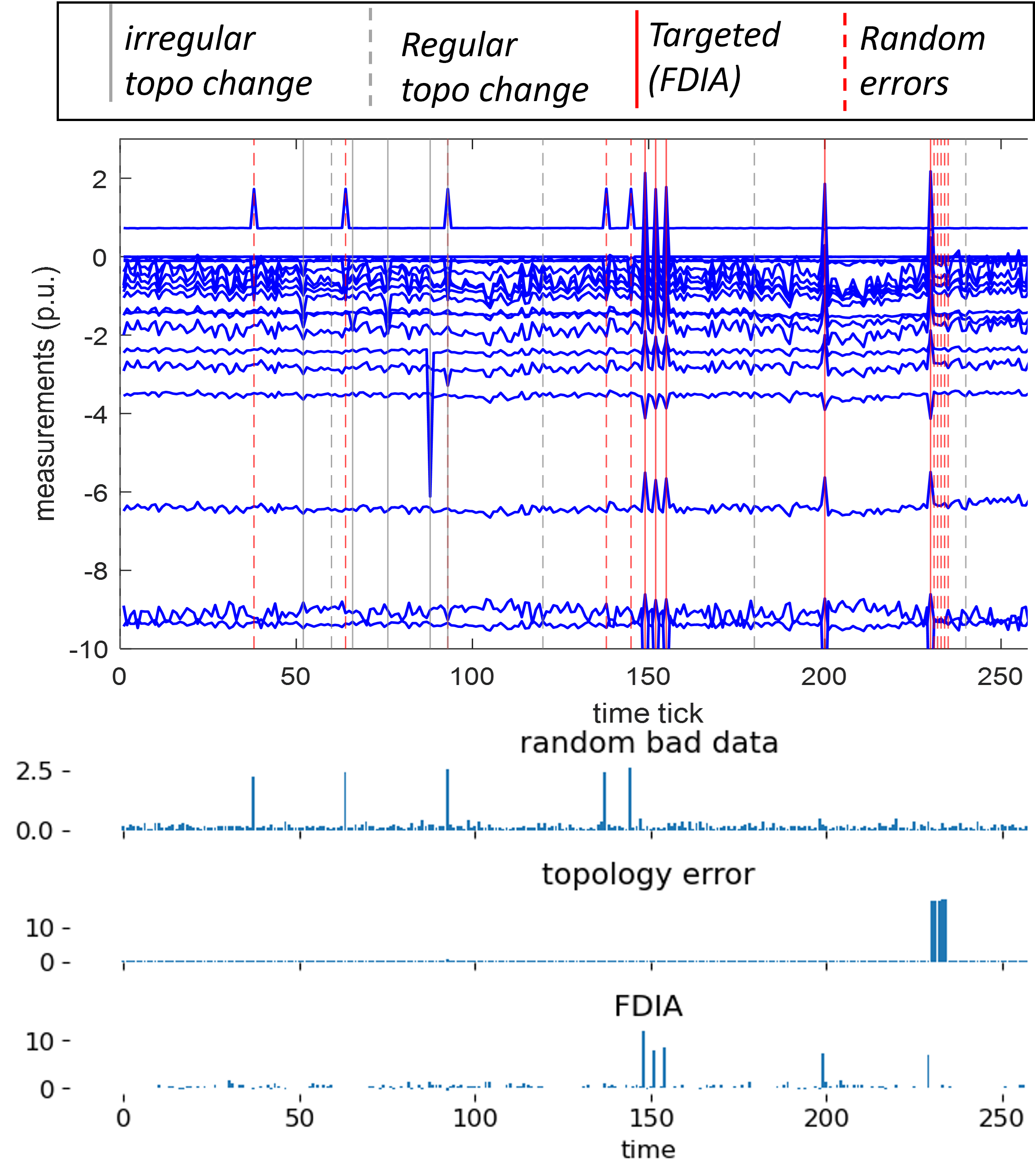}
     \caption{Anomaly detection on 1354-bus system.}
     \label{fig: 1354pegase AD}
     \end{figure}
     
\begin{figure*}[!t]
    \centering
    \includegraphics[width=0.95\linewidth]{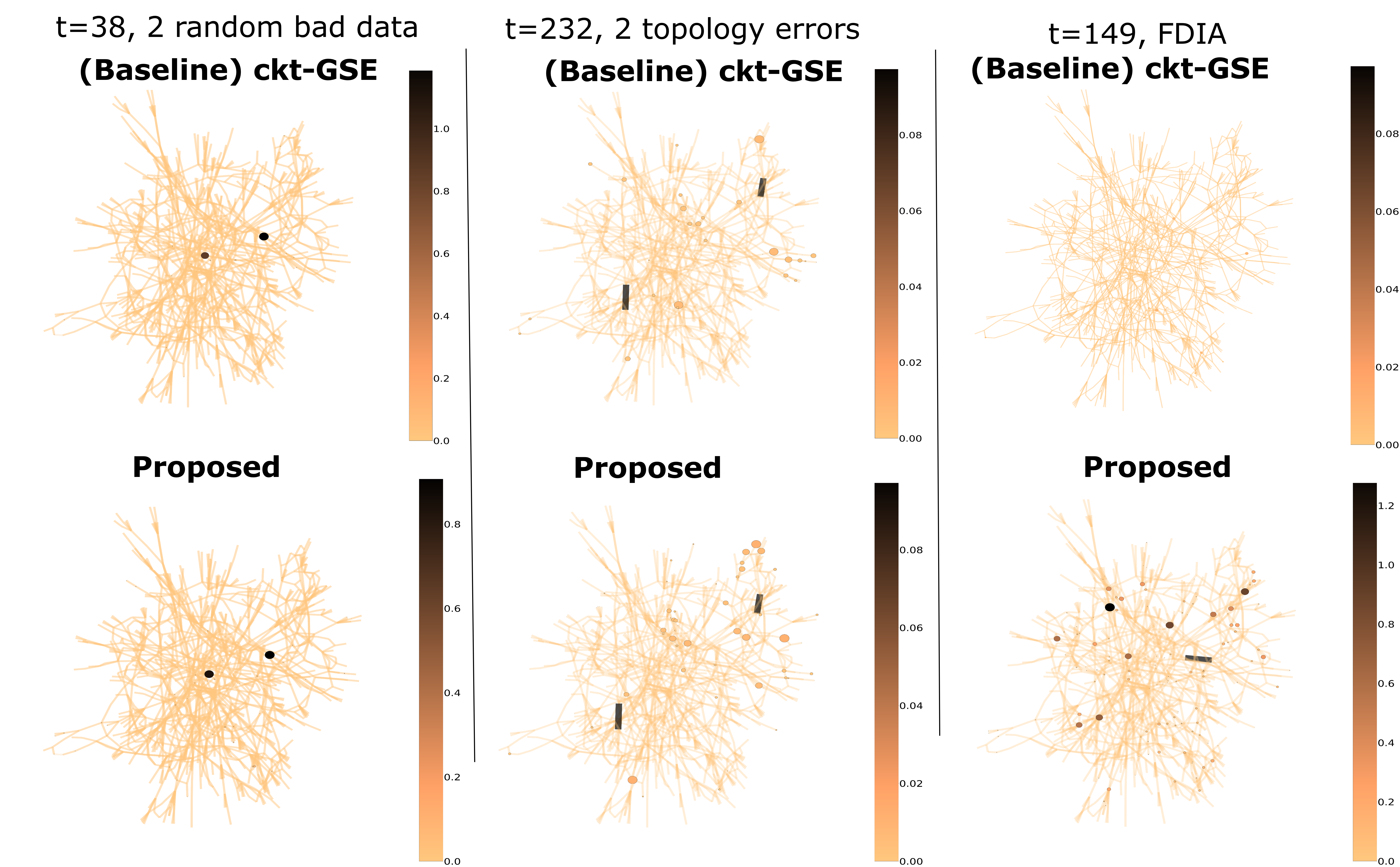}
    \caption{Anomaly localization on 1354-bus system.}
    \label{fig: 1354pegase, anomaly localization}
\end{figure*}

\end{document}